\documentclass[10pt,a4paper]{article}
\pdfoutput=1

\usepackage{microtype}
\DisableLigatures[f]{encoding = *, family = * }

\usepackage{graphicx}

\usepackage{cite}
\usepackage{amsmath}

\begin{document}
%
\title{On the Use of a Spectral Glottal Model\\ for the Source-filter Separation of Speech}
%
%
%

\author{Olivier~Perrotin
        and~Ian~V.~McLoughlin, {Senior member,~IEEE}
\thanks{O. Perrotin and I.V. McLoughlin are with the School
of Computing, University of Kent, Medway, UK, e-mail: ofmp@kent.ac.uk}
}

\maketitle

\begin{abstract}
The estimation of glottal flow from a speech waveform is a key method for speech analysis and parameterization. Significant research effort has been made to dissociate the first vocal tract resonance from the glottal formant (the low-frequency resonance describing the open-phase of the vocal fold vibration). However few methods cope with estimation of high-frequency spectral tilt to describe the return-phase of the vocal fold vibration, which is crucial to the perception of vocal effort. This paper proposes an improved version of the well-known Iterative Adaptive Inverse Filtering (IAIF) called GFM-IAIF. GFM-IAIF includes a full spectral model of the glottis that incorporates both glottal formant and spectral tilt features. Comparisons with the standard IAIF method show that while GFM-IAIF maintains good performance on vocal tract removal, it significantly improves the perceptive timbral variations associated to vocal effort.
\end{abstract}


%

\section{Introduction}
%
%
%
%
Speech communication is the combination of a linguistic component which conveys messages through the articulation of phonemes, and a prosodic component which encodes speech expression through variations of pitch, intensity, rhythm and timbre.
The widely used linear model of source production \cite{Fan70} models those components independently, in 
four parts; an excitation combining pulse train and with noise to convey information on pitch, intensity, and breathiness; a glottis filter modeling the vibration shape of the vocal folds to convey information on voice quality (i.e. timbre); a vocal tract (VT) filter modeling the oral and nasal cavity resonances responsible for the perception of phonemes; and a lip radiation filter which mainly has the effect of a derivative filter. In the frequency domain the speech is then computed as $S(\omega) = E(\omega) G(\omega) V(\omega) L(\omega)$, where
$S, E$ are the spectra of the speech and excitation, and $G, V, L$ are the frequency responses of the glottis, VT, and lip radiation filters, respectively. 
$G$ and $L$ are often combined to provide a glottal flow derivative rather than a glottal flow.

A typical non-intrusive approach to study each component is glottal inverse filtering (GIF) \cite{Dru14a}. This relies on accurate estimation of the glottis and VT filters from the acoustic signal, and deconvolves the VT filter from the latter.  Unfortunately, while the spectral characteristics of the glottis are broadband, most current methods associate the lower part of the spectrum with the glottis and the higher part with the VT. High-frequency glottis features are thus generally assigned to $V$ rather than $G$. 
While this approximation works for applications involving moderate speech, it breaks down at the extremes of voice quality where high frequency variations reflect glottal behaviour changes due to vocal effort.
This letter proposes an improved version of the well-known Iterative Adaptive Inverse Filtering (IAIF) method for GIF, to extract the full spectral characteristics of the glottis filter.

\section{Review of glottal inverse filtering methods}
\label{sec:review} 

\subsection{Spectral model of voice production}

\subsubsection{Glottis}

\begin{figure}[!t]
\centering
\includegraphics[width=0.65\linewidth]{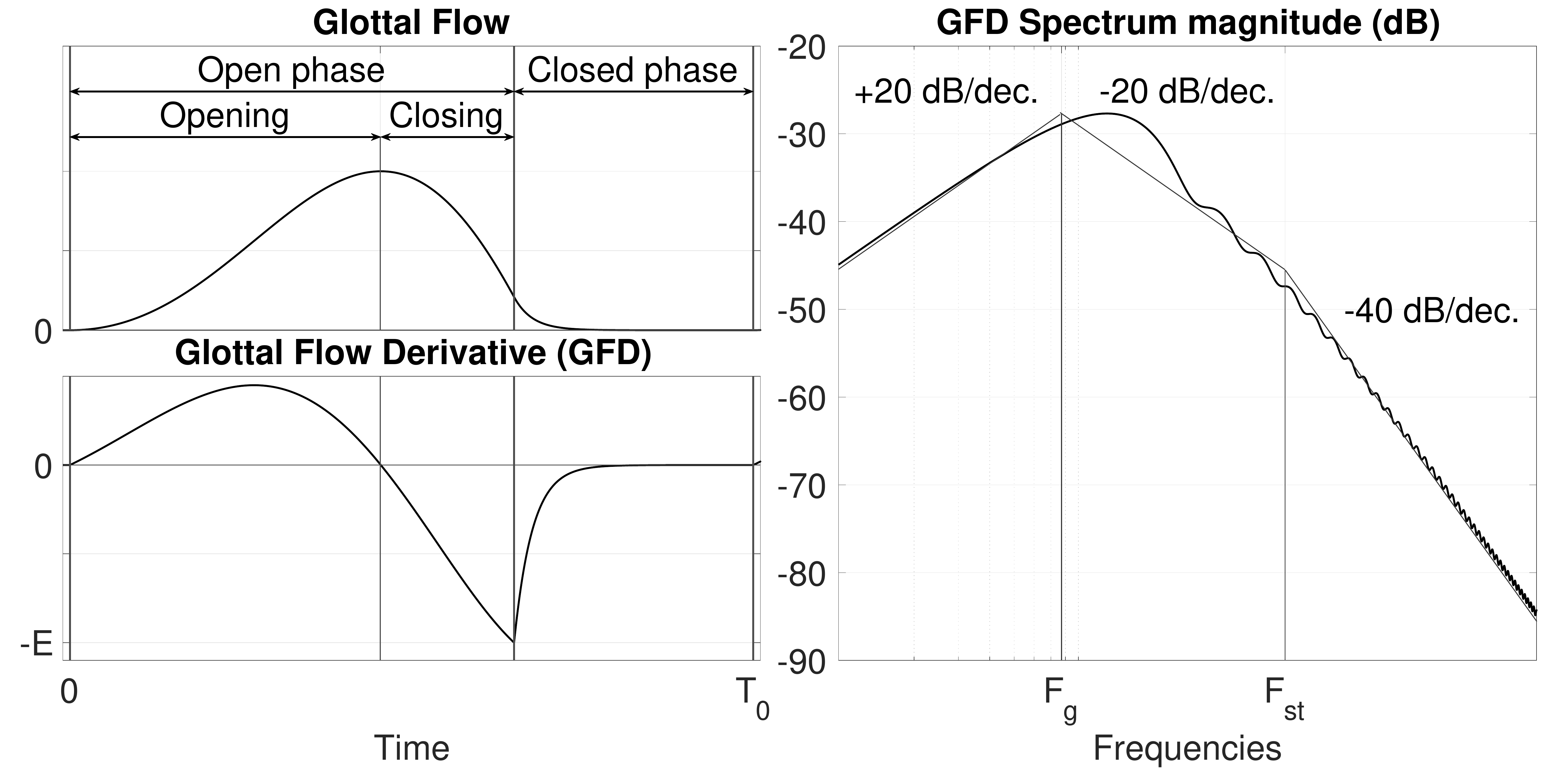}
\caption{LF-model. Top-left: Glottal flow model; Bottom-left: Glottal flow derivative model; Right: Glottal flow derivative spectrum}
\label{fig:LFmodel}
\end{figure}

One period of vocal fold vibration starts with an opening phase, where the folds are pulled apart by sub-glottal pressure. When the pressure becomes weaker than the elasticity of the vocal folds, the latter are drawn closer to eventually close the trachea aperture. This is the closing phase. The folds then remain closed until the sub-glottal pressure becomes large enough to trigger a new opening phase. The left part of Fig.~\ref{fig:LFmodel} depicts one vocal flow period (top) and its derivative (bottom) from the widely used LF-model~\cite{Fan85}.

It has been shown that open and closed phases are represented by distinct regions of the frequency spectrum~\cite{Dov06}. In particular, the oscillation provoked by the open phase leads to a major peak near the fundamental frequency often called the ``glottal formant". This is easily modeled by a second-order all-pole resonant filter with a $\pm$20~dB/decade slope (see right panel of Fig.~\ref{fig:LFmodel}). 
The position $F_g$ and bandwidth $B_g$ of the glottal formant are linked to the relative duration of the open phase over a period and, the glottal pulse asymmetry~\cite{Fan95,Hen01b}.

The abruptness of the closing phase relates to the high frequencies of the spectrum. The smoother the closure, the lower the cutting frequency $F_{st}$ of the spectral tilt, and the more attenuated the high frequencies. 
A supplementary $-20$\,dB/decade first order low-pass filter accounts for this.
In summary, the glottal flow derivative can be modeled by a third order filter with a complex conjugate pole pair $\{a,a^{*}\}$ (glottal formant) and one real pole $b$ (spectral tilt):
\begin{equation}
\label{eq:G}
G(z) = \frac{1}{ (1 - az^{-1}) (1 - a^{*}z^{-1}) (1 - bz^{-1}) }
\end{equation}

Many studies describe voice quality through this model.
For instance, a tensed voice has a higher and wider glottal formant and smaller spectral tilt than a breathy voice~\cite{Chi91}. 
A close correlation has also been found between vocal effort and spectral tilt~\cite{Har11,Duv13}. 
These results motivate the use of the \mbox{3-pole} glottal flow model for voice quality modification~\cite{dAl98b,Per16c} and expressive singing or speech synthesis~\cite{Feu17,Gob03}.

\subsubsection{Vocal tract} The oral cavity introduces resonances (poles) in the glottal flow spectrum while the nasal cavity attenuates certain regions (zeros). Nevertheless, the VT is often simplified by neglecting the nasal contribution (or by approximating it with a pair of additional resonances~\cite{Osh88}), and expressed as an auto-regressive model composed of $N_v$ pairs of complex conjugate poles $\{c_i,{c_i}^{*}\}$:
\begin{equation}
V(z) = \frac{1}{ \prod_{i=1}^{N_v} (1 - {c_i}z^{-1}) (1 - {c_i}^{*}z^{-1}) }
\end{equation}

\subsubsection{Lip radiation} Airflow radiation at the lips is often modeled as a derivative filter with coefficient $d$ close to 1~\cite{CUPbook2016}, $L(z) = 1 - d z^{-1}$

\subsection{Glottal inverse filtering methods}

Glottal inverse filtering has been investigated for the past 60 years \cite{Deg10}, \cite{Alk11} with the most straightforward methods using linear prediction to extract the VT after pre-emphasis of the speech signal. 
Assuming that the glottis filter can be reduced to the glottal formant contribution and that the $a$ coefficient in the $G$ filter is close to the $d$ coefficient in the $L$ filter, the contribution of $GL$ can be removed by a simple first-order high-pass filter~\cite{Dov97}. 
IAIF uses $1^{st}$ order LPC analysis to define the pre-emphasis filter \cite{Alk92}, although more thorough pre-emphasis filter estimation has been proposed~\cite{Aka05}.
Another way to remove the effect of the glottis is to apply linear prediction to the VT during the closed-phase \cite{Won79}. 
This requires accurate detection of the closed phase instants.

Mixed-phase decomposition assumes that the glottal formant is anticausal ($|a| > 1$) and that the spectral tilt and VT are causal ($|b| < 1$ and $|c| < 1$). Therefore, separation of the minimum phase and maximum phase components through the zeros of the Z transform \cite{Boz05}, or the complex cepstrum of speech \cite{Dru11}, leads to the extraction of the glottal formant on one hand, and the VT plus spectral tilt on the other hand.
Both pre-emphasized linear prediction and mixed-phase decomposition model the source as only a glottal formant. Therefore the spectral tilt remains in the VT filter and the GIF is not exhaustive.

To model the full glottis in the GIF process, other techniques use a glottal model in the estimation process. Joint optimization of VT and glottal models has been suggested~\cite{Hed84,Vin07}, considering speech as an autoregressive model with parametric glottal excitation. Although the estimation quality is high, the technique has high computational complexity and suffers from convergence issues.
Finally, by combining pre-emphasis with a glottal model, some methods estimate a glottal model from the signal, deconvolve the model to the signal, and apply linear prediction to estimate the VT. Glottal model codebooks have been proposed~\cite{Shu09} while other authors estimate glottis parameters directly from the signal~\cite{Cab13,Deg13a}. Optimization techniques are required to estimate the glottis parameters.

IAIF is probably the most popular method, combining straightforward computation (no estimation or optimization needed) with no requirement for \textit{a priori} knowledge of the signal. It is also noise robust, thus suitable for low-quality recordings \cite{Dru12b}. Despite this popularity, it does not encompass the spectral tilt of the glottis filter, which is important for conveying the perception of vocal effort. 
A recent attempt to encompass spectral tilt has been proposed in the IOP-IAIF method~\cite{Mok17}. This uses unconstrained high-order filtering for signal pre-emphasis and, when evaluated for a spoken /a/ at various levels of vocal effort, improved separation of voice qualities. 
We believe that while the extension of IAIF is merited, the unconstrained filter endows the glottal model with too much complexity, and is not straightforward to implement. 
Instead, we propose extending the first order glottal model of IAIF to a third order filter based on evidence that third order models are sufficient \cite{Chi91,dAl98b,Per16c,Feu17}. The benefits of a third order spectral glottal model in eqn. \ref{eq:G} are twofold: (1) we will demonstrate that it is significantly better than IAIF at conveying vocal effort and (2) it enables the extraction and modification of simple spectral parameters (e.g. $F_g$, $B_g$ and $F_{st}$), essential for voice transformation and synthesis.

\section{Glottal inverse filtering}
\label{sec:system}


We propose Iterative Adaptive Inverse Filtering with Glottal Flow Model (GFM-IAIF) to replace the simple IAIF model pre-emphasis filter with a $3^{rd}$ order glottal model. 
Traditional IAIF \cite{Alk92} is accomplished in four steps:
\begin{enumerate}
\item Gross glottis and lip filters estimation (1\textsuperscript{st} order LPC).
\item a. Remove glottis and lip filters from speech signal.\\b. Gross VT estimation (high order LPC).
\item a. Remove VT and lip filters from speech signal.\\b. Fine estimation of the glottis (high order LPC).
\item a. Remove glottis and lip filters from speech signal.\\b. Fine estimation of the VT (high order LPC). 
\end{enumerate}

Step (1) is critical to give the global shape to the glottis. Step (2) then encompasses the remaining spectral variations within the VT filter. Although the order of the fine estimation of the glottis filter is high, the estimated shape of the glottis spectrum will be globally close to that in step (1).
As stated above, IAIF assumes that the glottis can be modeled as the glottal formant only, by modeling the glottis and lip radiation filters as a first order filter.\\


\begin{figure}[!t]
\centering
\includegraphics[width=0.66\linewidth]{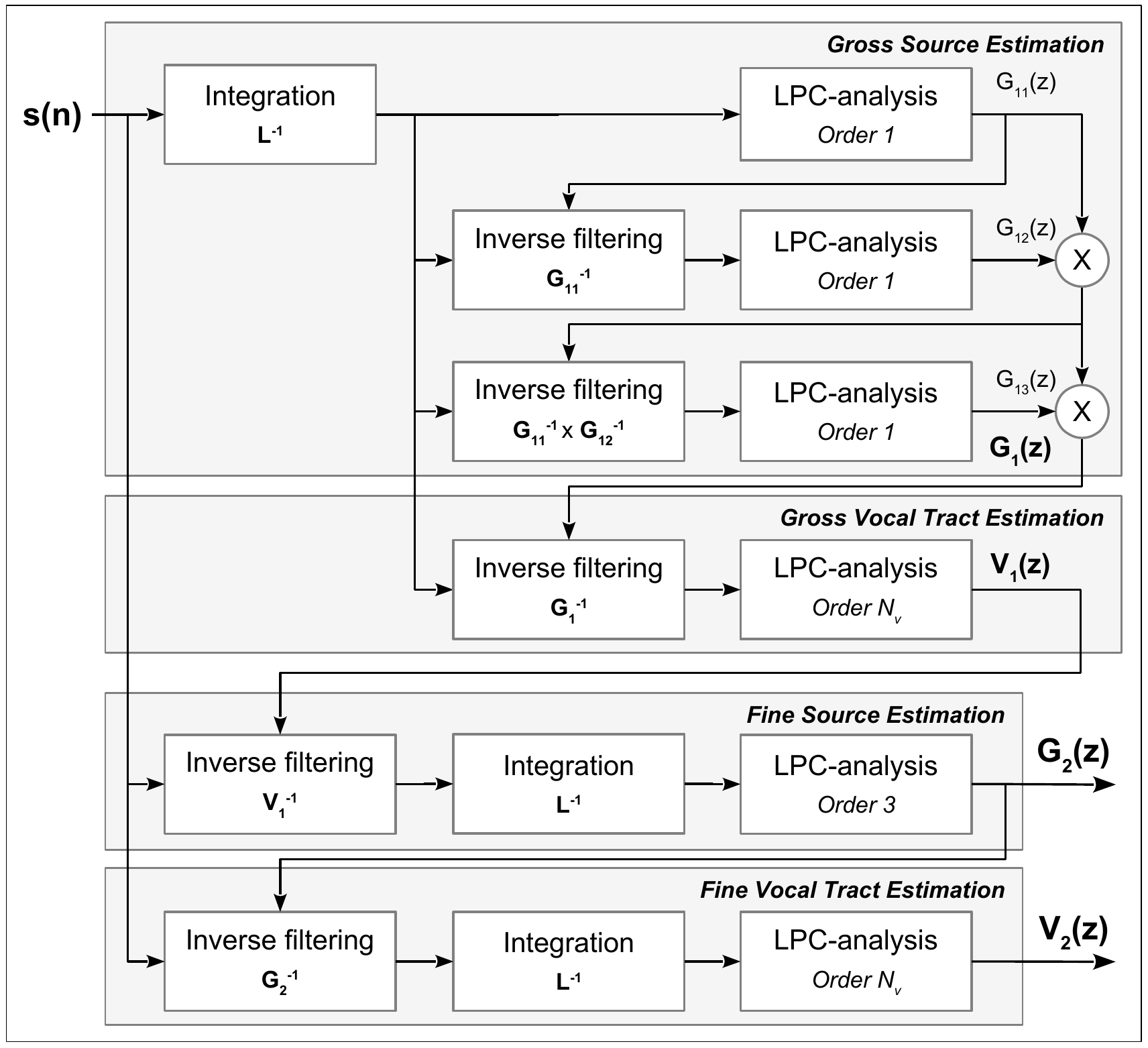}
\caption{Flowchart of the GFM-IAIF method}
\label{fig:flowchart}
\end{figure}

GFM-IAIF is proposed as shown in Fig.~\ref{fig:flowchart}. The architecture extends on IAIF by estimating the glottal flow (eqn.~\ref{eq:G}) during the gross estimation process as a $3^{rd}$ order filter. 
Moreover, it is essential during gross estimation of the glottis not to model any VT formants. 
For this sake, the estimation is accomplished by three successive first order iterations. 
While IAIF jointly estimated the glottis and lip radiation filters, GFM-IAIF first integrates the signal to remove the lip radiation contribution, then estimates the glottis independently. 
The resulting glottis filter has three real poles, and its frequency response can be stylized with slopes of $0$, $-20$, $-40$ and $-60$ dB/decade. In practice, the two first poles tend to have cutting frequencies close to each other at the position of the glottal formant.

The VT gross estimation phase follows IAIF: the gross glottis and lip radiation filters are deconvolved from the original signal and VT autoregressive coefficients then estimated through high order linear prediction. A third order LPC is then used during the fine estimation of the glottis to ensure that the final glottis filter follows equation~\ref{eq:G}.
The VT fine estimation is identical to the method in IAIF.

\section{Evaluation}
\label{sec:evaluation}

GIF evaluation is complicated by the lack of glottal flow ground truth. However, as the glottal source is expected to convey voice quality information~\cite{Dal06a}, Drugman \textit{et. al} evaluated GIF based on its ability to encode voice quality. 
The better the GIF method, the better voice quality is conveyed~\cite{Dru12b}.
We will use the same database and criteria as the above authors to evaluate GFM-IAIF,
namely, 12 different vowels uttered at three vocal effort levels (soft, medium, loud) by a single German female speaker from the de7 diphone database \cite{de7,Sch03}.
Each vowel was repeated from 18 to 23 times with the same vocal effort, leading to 825 stimuli, sampled at 22.05 kHz.

IAIF (using the COVAREP toolbox~\cite{Deg14}), the recently published IOP-IAIF \cite{Mok17} and GFM-IAIF methods are compared. 
Each method uses the IAIF default parameters of lip coefficient $d = 0.99$ and VT LPC order $N_v = F_s/1000 +4 = 26$. The glottis LPC order for fine estimation was set to $N_g = 3$ to match the spectral glottal model, for all methods.


\subsection{Glottal inverse filtering}

\begin{figure*}[!tp]
\centering
\includegraphics[width=\linewidth]{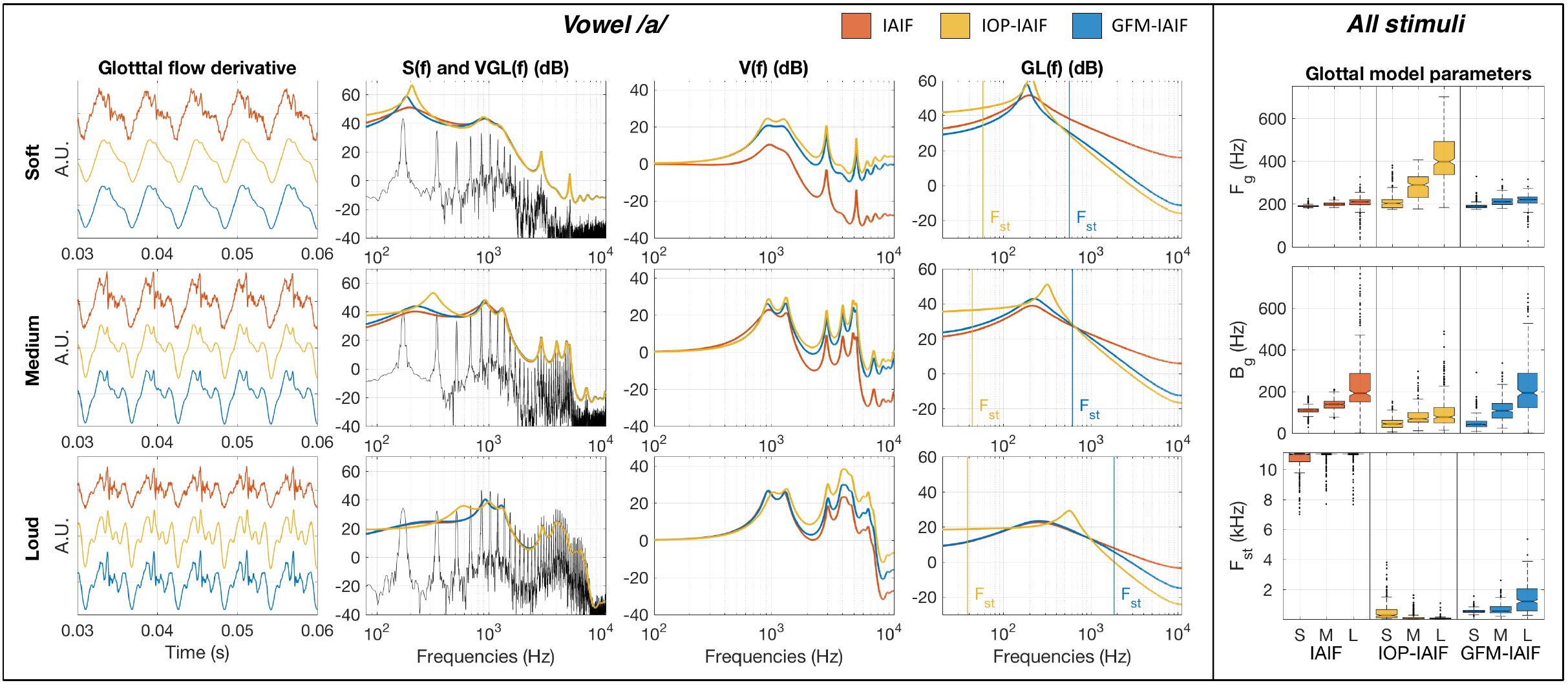}
\caption{Plots of decomposed signals for IAIF (orange), IOP-IAIF (yellow) and GFM-IAIF (blue) for vowel /a/ for soft (S): top row; medium (M): middle row and loud (L): bottom row stimuli. The right-hand column plots the distribution of the glottal formant frequency (top); glottal formant bandwidth (middle) and spectral tilt cutting frequency (bottom) for all stimuli depending on the method and voice qualities.}
\label{fig:extraction}
\end{figure*}

Fig.~\ref{fig:extraction} explores the three GIF methods for vowel /a/ uttered with the three voice qualities. Glottal flow derivatives all exhibit some ripples resulting from incomplete formant extraction, commonly observed with IAIF~\cite{Dru12b,Stu07}. Nevertheless, IOP-IAIF and GFM-IAIF show better variability across the three voice qualities.

Glottis extraction (4\textsuperscript{th} column) shows that while GFM-IAIF exhibits stronger variations of bandwidth with a narrower glottal formant for soft voice than IAIF, IOP-IAIF provides extremely high glottal formant positions and low bandwidths for all voice qualities. Moreover, the spectral tilt is not equally balanced between glottis and VT spectral envelopes across methods. The glottal spectral tilt for IAIF is low and constant for all voice qualities, resulting in high tilt variations in the VT. Conversely, GFM-IAIF assigns most of the spectral tilt variations in the glottis. The spectral tilt cutting frequency increases for louder voices, leading to less tilt variation in the VT. As for the IOP-IAIF method, it seems the spectral tilt is maximum for all voice qualities, causing larger tilt variations in the VT envelope compared to GFM-IAIF.

The distributions of $F_g, B_g$ and $F_{st}$ displayed on Fig.~\ref{fig:extraction} (right column) suggest these trends are valid for all stimuli. A Wilcoxon rank-sum test assessed the difference between distributions relatively to voice quality (soft vs. medium; medium vs. large; small vs. large). All pairs were significantly different ($p < 10^{-3}$) except for the medium vs. loud distributions of $B_g$ extracted by IOP-IAIF, and the small vs. medium distributions of $F_{st}$ extracted by GFM-IAIF.
To summarize, GFM-IAIF features glottal formant parameters with the same order of magnitude as IAIF, with greater variability between voice qualities and in line with the literature~\cite{Dov06}. IOP-IAIF provides unexpectedly high formant positions and low bandwidths for medium/loud voices. Additionally, only GFM-IAIF provides the expected relationship between spectral tilt cutting frequency and voice quality \cite{Duv13}.

The main difference between IOP-IAIF and GFM-IAIF is the order of the gross estimation of the glottis. This is unconstrained in the former, reaching an LPC order of 19 for some of these stimuli. When restrained to a lower third order during fine estimation, IOP-IAIF encompasses the maximum slope in the glottis by keeping the spectral tilt cutting frequency low and narrowing the formant bandwidth.

\subsection{Voice quality classification}

\begin{figure}[!tb]
\centering
\includegraphics[width=0.65\linewidth]{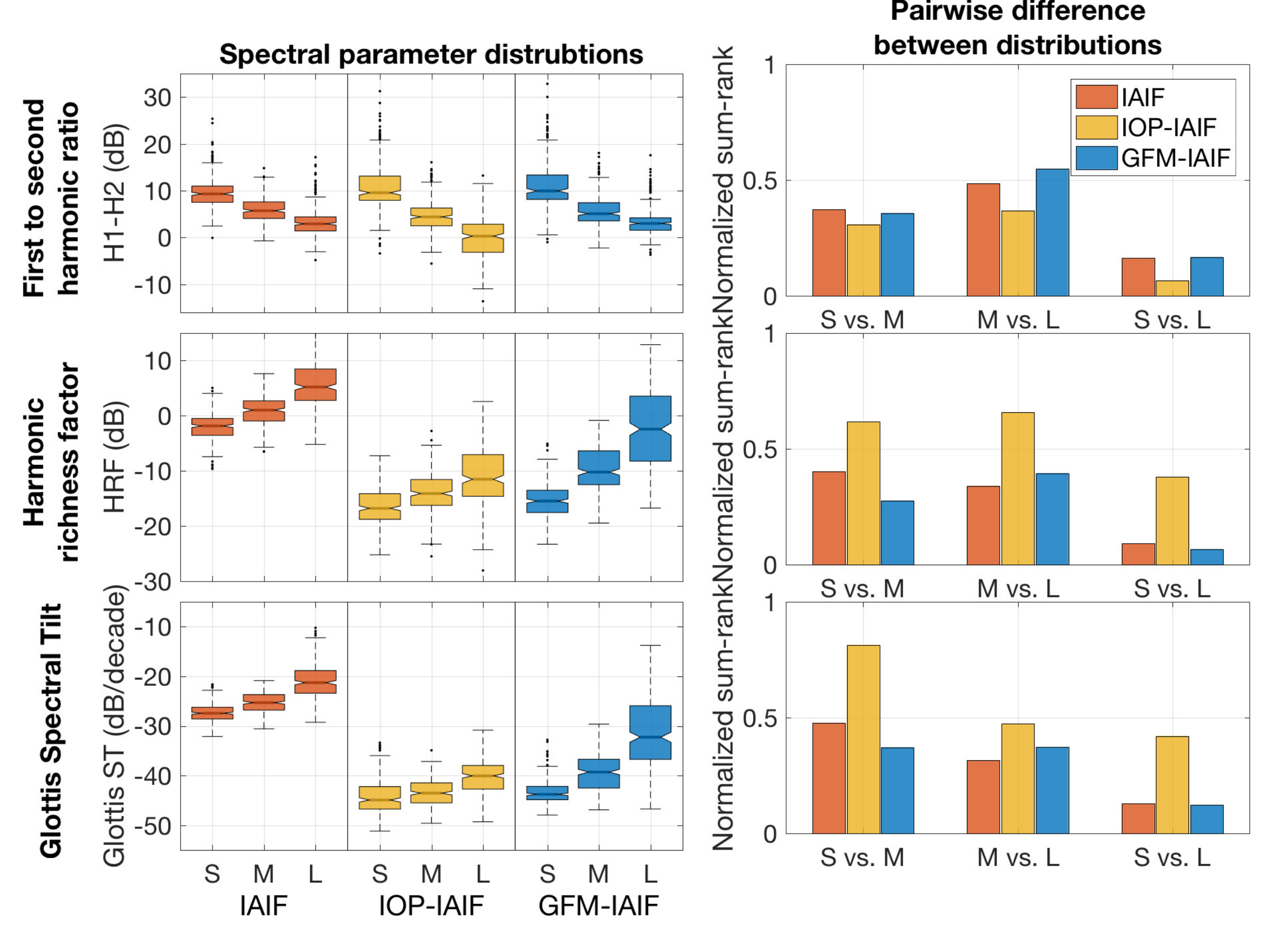}
\caption{Spectral parameters (top: H1-H2; middle: HRF; bottom: glottis ST). Left: distributions by extraction method and voice quality (S: soft; M: medium; L: loud). Right: normalized distribution rank-sum compared pairwise by voice quality for each parameter and method.}
\label{fig:specParam}
\end{figure}

We have seen how GFM-IAIF leads to glottal parameter variations that are more representative of voice quality, but is voice quality more predictable from the glottal flow derivative in GFM-IAIF than in other methods?
Glottal flow derivative can be described by several frequency-domain features \cite{Dru12b}.
The amplitude difference between first and second harmonics \textit{H1-H2} is linked to the glottal formant position \cite{Fan95,Kla90}; The closer it is to the first harmonic, the higher is \textit{H1-H2}. The harmonic richness factor \textit{HRF} is a measure of the quantity of harmonics in the spectrum, defined as the ratio between the sum of the $2^{nd}$ to $n^{th}$ harmonic amplitudes (in dB) over the fundamental frequency amplitude \cite{Chi91}. 
Finally, spectral tilt \textit{ST} (in dB/decade) is computed from a linear regression of the $n$ first harmonic amplitudes on a log-frequency scale. We chose $n$ to select harmonics below 5\,kHz only~\cite{Deg14}.
 As the voice becomes louder, smaller \textit{H1-H2} and higher \textit{HRF} and \textit{ST} values are expected ~\cite{Fan95, Chi91,Duv13}.

Fig.~\ref{fig:specParam} displays the distribution of these three parameters (top to bottom left) depending on voice quality and estimation method. We actually observe decreasing \textit{H1-H2} and increasing \textit{HRF} and \textit{ST} values with increasing vocal effort. 
A non-parametric Wilcoxon rank-sum test assessed the significance between different pairs of distributions (soft vs. medium; medium vs. large; small vs. large). Given the large sample size, all differences were assessed significant ($p < 10^{-3}$). 

The rank-sum calculated from each pair was normalized and displayed in Fig. \ref{fig:specParam} (right column). $0$ indicates non-overlapping distributions while $1$ indicates similarity. Hence, lower values denote more distinct distributions and greater likelihood the parameter could be used for voice quality discrimination. \textit{H1-H2} reflects the behaviour of $F_g$ observed previously. As IOP-IAIF provides higher variations of $F_g$, this affects \textit{H1-H2}, giving more distinct distributions across voice qualities than for IAIF or GFM-IAIF and lower normalized rank-sum values. However, as $F_g$ values were unexpectedly high for IOP-IAIF, one may question the relevance of \textit{H1-H2} as a discriminative parameter for this method.
IAIF and GFM-IAIF show similar performance and a good discriminative power for soft vs. loud voice qualities.
The lack of formant bandwidth variation with IOP-IAIF leads to the poorest \textit{HRF} discriminative performance.
GFM-IAIF shows the best discriminative power between soft vs. medium, and soft vs. loud. Moreover, the GFM-IAIF \textit{HRF} distributions are more spread than for IAIF.
GFM-IAIF \textit{ST} distributions also show better spread with voice quality. Nevertheless, IAIF shows as much discriminative power as GFM-IAIF. Although IAIF did not provide a variation of spectral tilt cutting frequencies, glottal formant bandwidth has an influence on \textit{ST} and justifies the discriminative performance of the latter. Finally, IOP-IAIF has the poorest performance as both glottal bandwidth and spectral tilt cutting frequency were wrongly detected.
Overall, it seems that GFM-IAIF can be slightly more discriminative than other methods regarding voice quality.

\section{Conclusion}
\label{sec:conclusion}

This paper proposes a new method for glottal inverse filtering, GFM-IAIF, having a third order filter in the pre-emphasis step, in line with spectral glottis source models. This models both the glottal formant and the spectral tilt effectively; two glottis spectral features responsible for the perception of voice quality.
Evaluation of GFM-IAIF against the standard IAIF and IOP-IAIF shows that GFM-IAIF provides the best estimation of glottal formant frequency and bandwidth, and spectral tilt cutting frequency depending on voice quality variations, according to literature \cite{Dov06,Chi91,Har11,Duv13}.

By ensuring a reduced set of parameters (frequency and bandwidth of glottal formant and cutting frequency of spectral tilt), GFM-IAIF also eases intuitive voice quality analysis and synthesis.


\end{document}